# Single-source, solvent-free, room temperature deposition of black γ-CsSnI$_3$ films

*Vivien M. Kiyek, Yorick A. Birkhölzer, Yury Smirnov, Martin Ledinsky, Zdenek Remes, Jamo Momand, Bart J. Kooi, Gertjan Koster, Guus Rijnders, Monica Morales-Masis\**

V.M. Kiyek, Y.A. Birkhölzer, Y. Smirnov, Prof. G. Koster, Prof. G. Rijnders, Dr. M. Morales-Masis

MESA+ Institute for Nanotechnology, University of Twente, P.O. Box 217, Enschede, 7500 AE, The Netherlands

Email: m.moralesmasis@utwente.nl

Dr. M. Ledinsky, Dr. Z. Remes

Institute of Physics, Academy of Sciences of the Czech Republic, Cukrovarnická 10, Prague, 162 00, Czech Republic

Dr. J. Momand, Prof. B.J. Kooi

Zernike Institute for Advanced Materials, University of Groningen, Nijenborgh 4, Groningen, 9747 AG, The Netherlands

**ABSTRACT:** The presence of a non-optically active polymorph (yellow-phase) competing with the optically active polymorph (black γ-phase) at room temperature in CsSnI$_3$ and the susceptibility of Sn to oxidation, represent two of the biggest obstacles for the exploitation of CsSnI$_3$ in optoelectronic devices. Here room-temperature single-source *in vacuum* deposition of smooth black γ–CsSnI$_3$ thin films is reported. This has been done by fabricating a solid target by completely solvent-free mixing of CsI and SnI$_2$ powders and isostatic pressing. By controlled laser ablation of the solid target on an arbitrary substrate at room temperature, the formation of CsSnI$_3$ thin films with optimal optical properties is demonstrated. The films present a band gap of 1.32 eV, a sharp absorption edge and near-infrared photoluminescence emission. These properties and X-ray diffraction of the thin films confirmed the formation of the orthorhombic (B-γ) perovskite phase. The thermal stability of the phase was ensured by applying *in situ* an $Al_2O_3$ capping layer. This work demonstrates the potential of pulsed laser deposition as a volatility-insensitive single-source growth technique of halide perovskites and represents a critical step forward in the development and future scalability of inorganic lead-free halide perovskites.

Pulsed Laser Deposition (PLD) has offered unique options for the development of complex materials thin film growth, allowing stoichiometric transfer and multi-compound deposition independent of the relative volatility of the elements and ultimate control of interfaces. In the field of complex oxides, PLD opened the way to high-$T_c$ superconducting films requiring stoichiometric transfer of multiple (4-5) cations.[1] Here we present the rather unexplored but enormous potential of PLD as a unique single-source in-vacuum deposition technique of all-inorganic halide perovskites, using $CsSnI_3$ as case example.

$CsSnI_3$ has been widely proposed in literature as a Pb-free and all-inorganic alternative to the archetypical hybrid halide solar cell absorber, $CH_3NH_3PbI_3$ ($MAPbI_3$). The replacement of toxic Pb with Sn is a natural choice due to their similar ionic radius and lower toxicity of Sn.[2] The replacement of the organic cation (e.g. $CH_3NH_3$) with Cs has been proposed to enhance the thermal stability of the material.[3] While the decomposition temperature of Cs-based halide perovskites is higher than the ones containing organic cations, the size of the $Cs^+$ cation is at the limit for stability of the perovskite structure, and therefore causing phase instabilities[4,5,6] between the optically active (black) perovskite phase and the non-optically active (yellow) non-perovskite phase. In $CsSnI_3$ these phases can coexist at room temperature. Black phase stabilization in all-inorganic perovskites is therefore critical to ensure their application in optoelectronic devices, and has been the subject of very recent work, focused on $CsPbI_3$ [7,8] and $CsSnI_3$ [9].

In terms of synthesis, solution-based processes are the most widely used techniques to fabricate these materials.[10–14] Concerns about the use of highly toxic solvents and complex device integration have recently motivated the investigation of solvent-free and vacuum-based thin film deposition processes.[11,15,16] Thermal co-evaporation has been the main in-vacuum technique that enabled high quality thin film formation of a family of halide perovskites and high-efficiency devices.[11,17,18] However, the need for multiple sources due to the different volatility of the constituent elements poses a limitation for the synthesis of multication-multihalide materials and their further upscaling. Steps towards achieving a single-source deposition have sporadically been reported for laser-based techniques. This includes resonant infrared matrix assisted pulsed laser evaporation (RIR-MAPLE)[19,20] and pulsed-laser deposition (PLD)[21–23] of $MAPbI_3$ and $CsPbBr_3$, but high material quality remains yet to be demonstrated. An approach gaining popularity is the mechano-chemical synthesis of halide perovskite powders and subsequent thin film formation following single-source vapor deposition (SSVD) of those powders.[24–26] However, SSVD might present hurdles on the exploration of a plethora of multication-multihalides and double-perovskites due to differing volatilities, i.e. off-stoichiometric transfer and/or sticking.

Here we present single-source room temperature PLD of $CsSnI_3$ thin films with excellent optical properties achieved by the formation of the orthorhombic black (B-γ) perovskite phase. This work introduces PLD as an enabling technology to achieve near-stoichiometric transfer of all-inorganic halide perovskites in vacuum, opening the path for controlled and reproducible growth as well as ease of integration in devices from PV to more complex architectures such as integrated photonics.

**Figure 1** summarizes the fabrication process of the solid PLD target. Equimolar amounts of CsI and $SnI_2$ source powders were mixed by ball-milling in an Ar-filled vessel. We note that the mixing is done only by rotation of the cylindrical vessel containing the powder and $ZrO_2$ balls, and therefore is different from the known mechano-chemical synthesis.[26] To ensure a uniform mixture of the powders, the mixing process was left running for three days. The enhanced uniformity of the Cs, Sn and I elemental distribution on the target with increasing mixing time was confirmed by Energy-Dispersive X-ray Spectroscopy (**Figure S1 and Table S1**). The mixed powders were then pressed into a 2 cm diameter disk-shaped pellet using an uniaxial press, and subsequently exposed to an isostatic pressure of 360 MPa using a hydraulic press. A similar procedure has been used for the fabrication of $Cs_2AgBiBr_6$ wafers.[27] The isostatic pressing allows for the formation of a compact and dense target (> 85% calculated density) as required for PLD. Therefore, no further sintering with heating was required. It is important to note that the pressed solid target does not react into the $CsSnI_3$ phase, as shown by the X-ray diffraction pattern in Figure 3a.

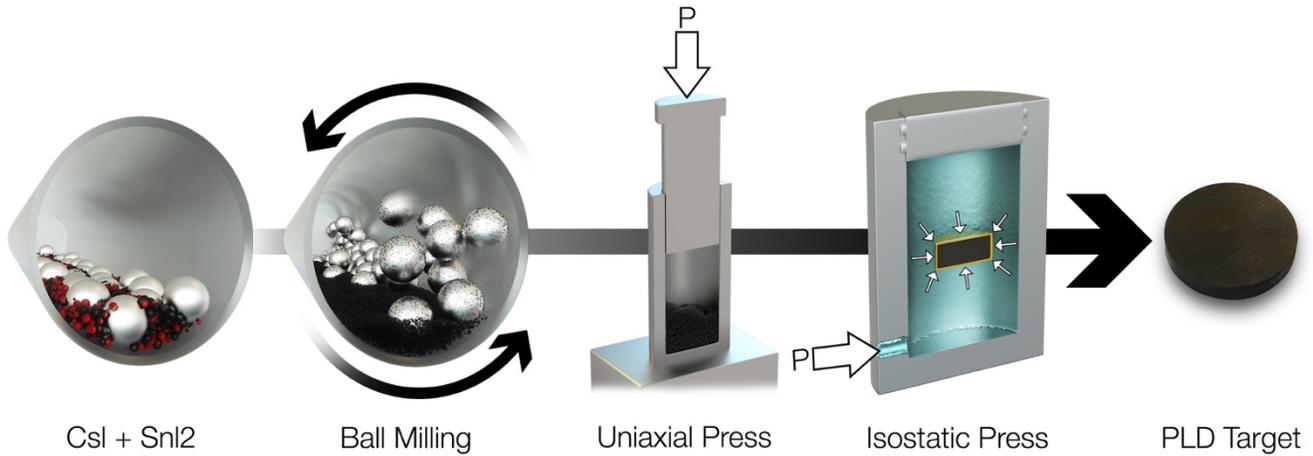

Figure 1. Illustration of PLD target fabrication process. From left to right: stoichiometric mixture of CsI and SnI$_2$ powders, ball milling, uniaxial press applying 33 MPa and hydraulic press applying 360 MPa isostatically, and final PLD solid target.

The target was loaded into the PLD chamber, which was then evacuated to a base pressure of ~1 × 10$^{-7}$ mbar. CsSnI$_3$ thin films were deposited onto Si (with native SiO$_x$), fused silica and glass substrates at room temperature using a KrF (248 nm) laser and a fluence of 0.2 J cm$^{-2}$. The growth rate was 0.05 nm/pulse such that for a laser repetition frequency of 5 Hz, the total duration to grow 100 (200) nm CsSnI3 was only 400 (800) seconds. An Ar working pressure of 1.3 × 10$^{-3}$ mbar was kept constant during deposition and no additional reactive gasses were introduced. Following CsSnI$_3$ deposition, an amorphous Al$_2$O$_3$ capping layer was applied *in situ* by PLD.

Steady-state photoluminescense (PL) spectra and absorption coefficient measurements were performed on 100 nm thick PLD-grown CsSnI$_3$ films on fused silica substrates (**Figure 2**). The PL emission is centered at 1.38 eV (900 nm). Consistently, the absorption coefficient determined by Photothermal Deflection Spectroscopy (PDS) shows a sharp absorption edge centered at 1.32 eV. The absorption coefficient is shown with a solid black line and for comparison, the absorption coefficient (also determined by PDS) of a reference methylammonium lead halide (MAPbI$_3$) perovskite film[28] is shown with a dashed black line. These results highlight the high absorption coefficient at the whole visible spectral range and sharp edge, indicating a high quality absorber material, comparable to MAPbI$_3$.[28] In order to extract the Urbach energy, the absorption spectrum was calculated from the PL spectrum at the band edge area via the reciprocal relation [29]. Individually, this recalculated absorption spectra and the absorption coefficient measured by PDS, both confirm an Urbach energy of 12.9 meV for the PLD grown

CsSnI$_3$ films. This is only 0.4 meV higher that of MAPbI$_3$ (12.5 meV) determined by the same methods. Such a low Urbach energy indicates potential for low voltage losses in the optimized solar cell.[30]

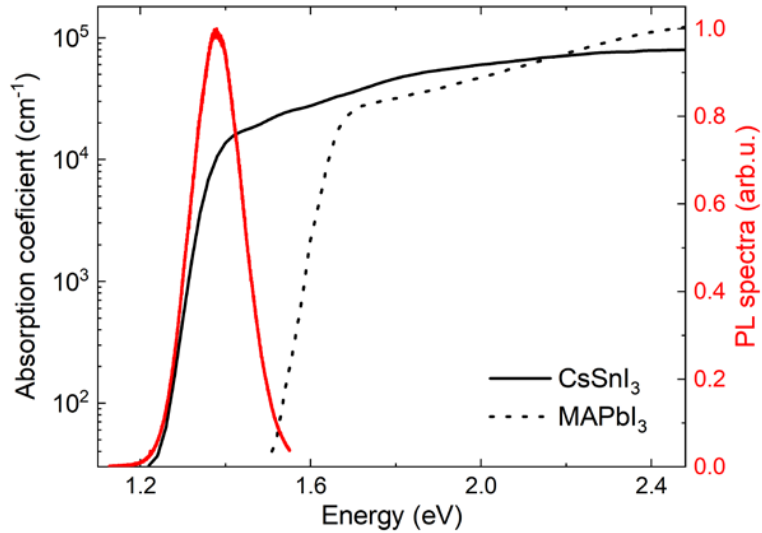

Figure 2. Absorption coefficient (left axis) and steady state PL (right axis) of 100 nm CsSnI3 PLD grown films. To highlight the sharp absorption edge of the CsSnI3, the absorption coefficient of a reference MAPbI3 film is presented.

A band gap of 1.32 eV extracted from the Tauc plot (**Figure S3**) and the aforementioned optical properties are characteristics of the black orthorhombic phase of CsSnI$_3$ (B-γ–CsSnI$_3$).[9] The formation of polycrystalline B-γ–CsSnI$_3$ thin films is confirmed by X-ray diffraction (XRD). **Figure 3** displays the XRD patterns of the target and the thin films shown in this manuscript. Panel (a) indicates that the target is an unreacted mixture of CsI and SnI$_2$ powders, whereas panel (b) demonstrates that thin films grown from this target by PLD at room temperature crystallize in the perovskite structure and very well match the reference pattern of B-γ–CsSnI$_3$ following reference.[9] No difference between thin (100 nm) and thicker films (200 nm) was noticed, indicating that the B-γ–CsSnI$_3$ phase remains stable with doubled thickness (solid lines in Figure 3b). The dashed lines in Figure 3b are measurements taken four months (~3000 h) after fabrication of the films, where the films were stored in a glove box filled with argon gas, demonstrating that the B-γ–CsSnI$_3$ phase of both film thicknesses remains stable even months after thin film fabrication. After these XRD measurements the same films were kept in open air for 72 hours, measured again, and still showed no structural changes (dotted lines in Figure 3b). The Al$_2$O$_3$ capping layer with a thickness of 13 and 40 nm for the 100 and 200 nm thick CsSnI$_3$ films, respectively, is amorphous and therefore not present in the XRD spectra. Information about thin films grown from targets with different mixing times is found in the supplementary information (**Figure S2**).

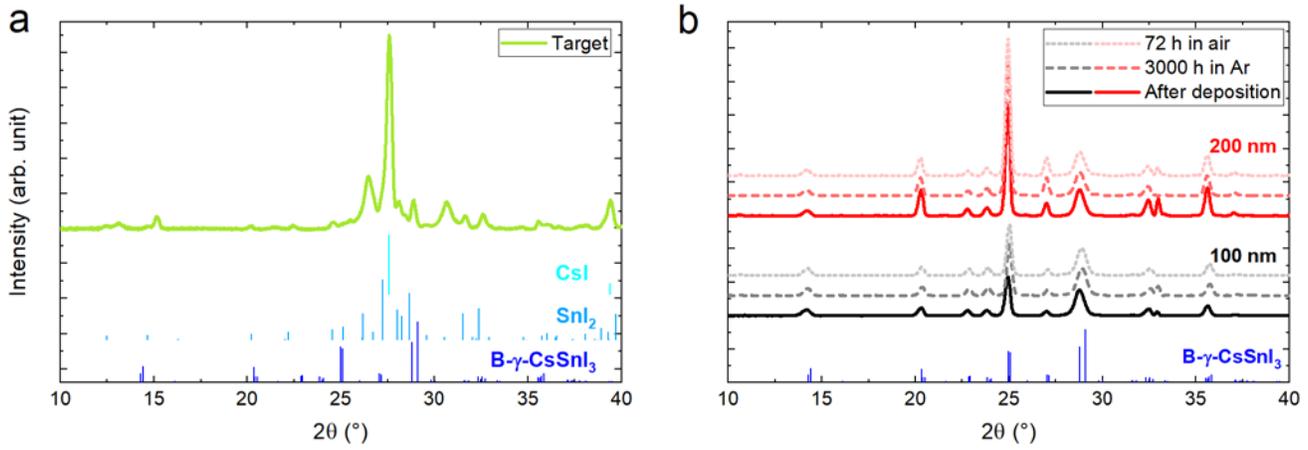

*Figure 3.(a) X-ray diffraction (XRD) pattern of the solid PLD target shown in Figure 1. (b) XRD patterns of 100 and 200 nm thick CsSnI$_3$ films deposited at room temperature by PLD. The solid lines are measurements directly after the deposition, the dashed lines after 3000 h in Ar atmosphere, and the dotted lines after additional 72 h in air. For comparison, the same B-γ-CsSnI$_3$ reference spectrum is plotted below in both graphs. The plots indicate that while the source target doesn't present the CsSnI$_3$ phase but a mixture of the CsI and SnI$_2$ powders, the resulting thin films present the single black orthorhombic phase of CsSnI$_3$, which is stable over long time thanks to the Al$_2$O$_3$ capping layer.*

Scanning Transmission Electron Microscopy (STEM) results (**Figure 4**) confirm uniform sticking of the ablated elements and the formation of a smooth and dense film with large elongated grains along the thickness of the film (Figure 4a). Zoomed-in area of Figure 4a with enhanced contrast is shown in Supporting Information, Figure S5. The distribution of cesium (Cs), tin (Sn) and iodide (I) in the films was evaluated by cross-section EDX mapping (Figure 4b-d). The EDX maps show a uniform distribution of the three elements across the thickness of the film and quantitative analysis indicate an overall Cs/Sn ratio of 1. Combining Rutherford Back Scattering (RBS) and Particle Induced X-ray Emission (PIXE), we determined an iodide content in the films of ~ 65 at% (**Figure S4**). Figure 4f-g also shows conformal coating and uniformity of the amorphous Al$_2$O$_3$ protective layer. The low roughness of the CsSnI$_3$ + capped Al$_2$O$_3$ films was furthermore confirmed by atomic force microscopy (supporting information Figure S6).

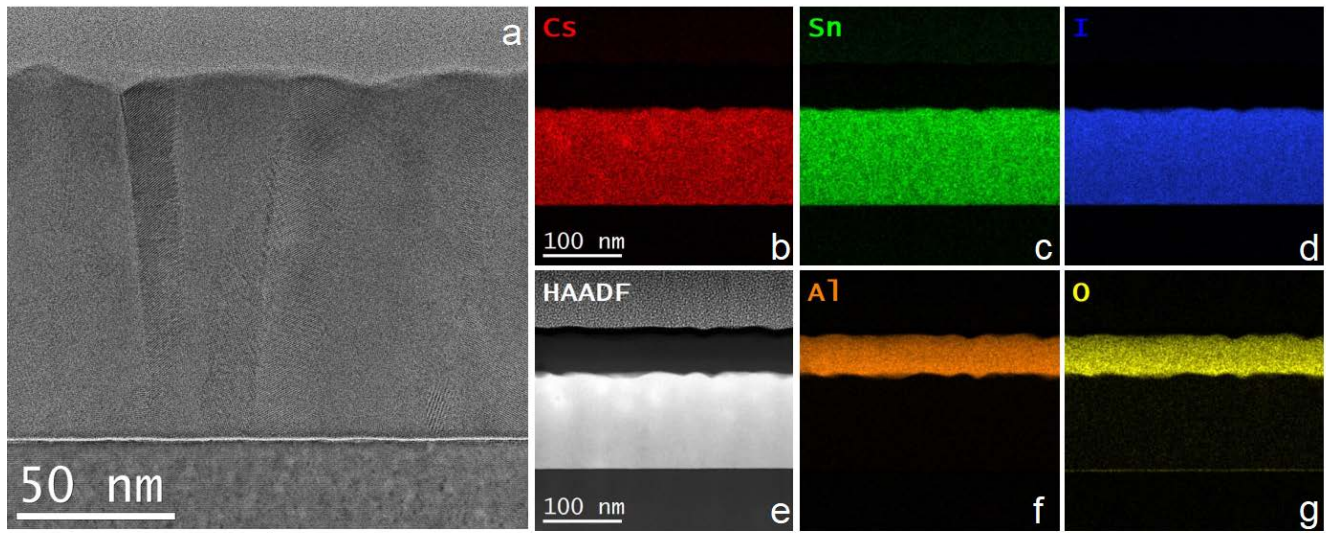

*Figure 4. Cross-section (a) bright-field TEM image of a PLD-grown B-γ-CsSnI₃ film on Si. The image shows the formation of a dense film with elongated crystalline grains. (b-g) High-angle annular dark-field (HAADF) image and EDS mapping of the constituent elements of CsSnI₃ and the Al₂O₃ capping layer. EDS confirms uniform distribution of Cs,Sn and I along the thickness of the layer and conformal coating of the Al₂O₃ capping layer.*

Concluding, we have demonstrated the feasibility of single-source in-vacuum deposition of B-γ–$CsSnI_3$ films by PLD. The high optical quality of the films and black phase confirmation by optical and structural characterization shows the enormous potential of PLD for the single source growth of halide perovskites even at room temperature. In comparison with recently reported SSVD, PLD presents the advantage of non-equilibrium ablation of a solid target, therefore allowing near-stoichiometric transfer insensitive to the different volatilities of the elements. Another demonstrated advantage of PLD is the possibility of depositing multilayers without breaking vacuum, in this case the application of the $Al_2O_3$ layer allowing the stabilization of the black phase and protection against oxidation. This work motivates further exploration of the electrical properties of the material as well as its integration in complex devices, such as absorbers in solar cell devices, monolithic tandem solar cells or efficient light emitters in integrated photonic circuits.

**Experimental Details**

*Methods and Materials*

Target source materials: CsI and SnI$_2$ source powders were purchased from Sigma-Aldrich (99.9% purity) and TCI (> 97.0% purity). For the Al$_2$O$_3$ deposition, an Al$_2$O$_3$ single crystal with rough surfaces to enhance laser absorption was used as source target.

PLD: The vacuum chamber was evacuated to a base pressure of ~1 × 10$^{-7}$ mbar. A KrF (248 nm) laser was used to ablate the fabricated CsSnI$_3$ target. Thin films were deposited onto Si (with native SiO$_x$), fused silica and glass substrates at room temperature. An Ar working pressure of 1.3 × 10$^{-3}$ mbar was kept constant during deposition and no additional reactive gasses were introduced. The laser frequency was kept at 5 Hz, target-to-substrate distance at 50 mm, and a fluence of 0.2 J cm$^{-2}$ was used. The deposition of the Al$_2$O$_3$ capping layer was performed under Ar atmosphere and room temperature as the CsSnI$_3$ film. Al$_2$O$_3$ is a large band gap material (~7 eV) with insignificant absorption in the measured spectral region, therefore, the measured optical properties are unaffected by the Al$_2$O$_3$ thin film.

PDS: Photothermal deflection spectroscopy directly measures the optical absorption of thin films with sensitivity of up to four orders of magnitude. The light absorption is determined via a sample heating effect, by measuring the deflection of a probe laser beam with a position-sensitive detector. The PDS spectrophotometer uses a 150 W Xe lamp as a light source and a monochromator equipped with grating blazed at 750 nm operating in a broad spectral range from ultraviolet to infrared region 400–1200 nm.

PL: Photoluminescence spectra are measured using an excitation laser at 442 nm in a Renishaw in-Via REFLEX spectroscope. The intensity of the excitation light is reduced in order to prevent any structural degradation during measurements.

XRD: The films were analyzed by X-ray diffraction, using a Bruker D8 Discover diffractometer with a high brilliance microfocus Cu rotating anode generator, Montel optics, a 1 mm pinhole beam collimator and an EIGER2 R 500K area detector.

TEM: Cross-sectional specimen were prepared with a FEI Helios G4 CX FIB, using gradually decreasing acceleration voltages of 30 kV, 5 kV and 2 kV. TEM analyses were performed with a double aberration corrected FEI Themis Z, operated at 300 kV. HAADF-STEM images were recorded with a probe currents between 50 and 200 pA, convergence semi-angle 21 mrad and HAADF collection angles 61-200 mrad. EDX spectrum imaging was performed with a probe current of 1 nA, where the spectra were recorded with a Dual-X system, providing in total 1.76 sr EDX detectors.

**Supporting Information**

Supporting Information is available from the Wiley Online Library or from the author.

**ACKNOWLEDGMENT**

The authors acknowledge Frank Roesthuis for support with the PLD system, Mark Smithers for high resolution scanning electron microscopy, and Max Döbeli for RBS & PIXE measurements. M.M.M. acknowledges the European Research Council (ERC) under the European Union's Horizon 2020 research and innovation programme (CREATE, Grant Agreement Number 852722), and UTWIST program of the University of Twente. M.L. acknowledges the support of Czech Science Foundation Project No. 17-26041Y.

**Conflict**
The authors declare no conflict of interest.

**Author Contributions**

V.M. Kiyek and Y.A. Birkhölzer contributed equally to this work.

**ORCID**

Vivien M. Kiyek: 0000-0001-9360-9541
Yorick A. Birkhölzer: 0000-0003-3133-2481
Yury Smirnov: 0000-0003-0417-4270
Martin Ledinsky: 0000-0002-6586-5473
Zdenek Remes: 0000-0002-3512-9256
Jamo Momand: 0000-0001-5153-2427
Bart J. Kooi: 0000-0002-0311-4105
Gertjan Koster: 0000-0001-5478-7329
Guus Rijnders: 0000-0002-4638-0630
Monica Morales-Masis: 0000-0003-0390-6839

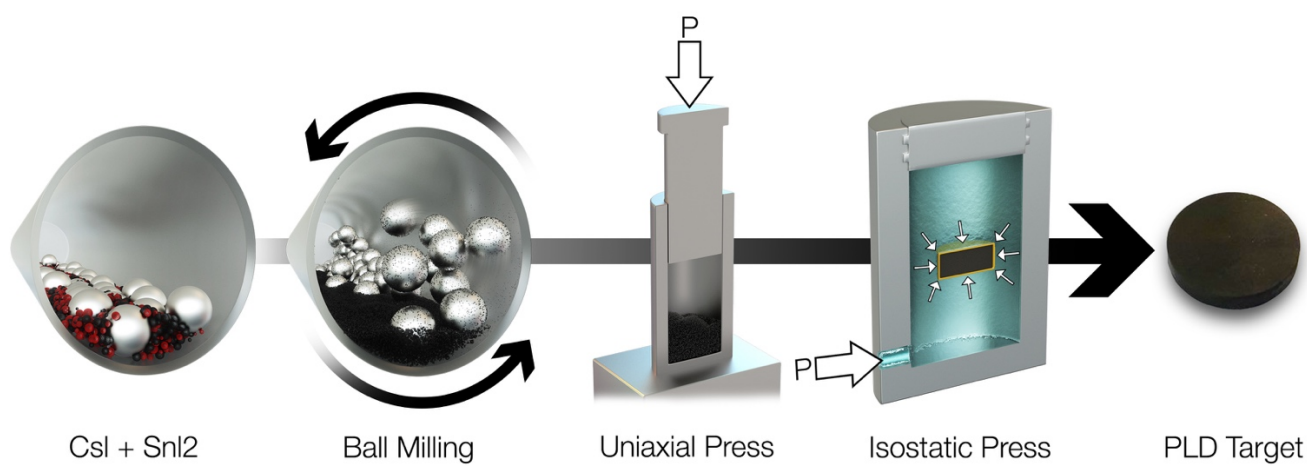

*Figure 1. Illustration of PLD target fabrication process. From left to right: stoichiometric mixture of CsI and SnI$_2$ powders, ball milling, uniaxial press applying 33 MPa and hydraulic press applying 360 MPa isostatically, and final target.*

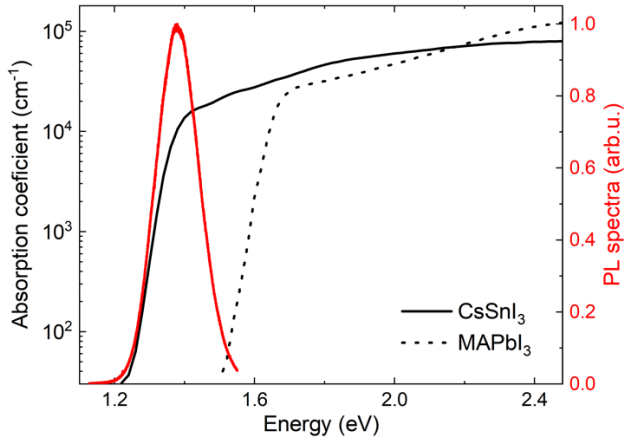

*Figure 2. Absorption coefficient (left axis) and steady state PL (right axis) of 100 nm CsSnI₃ PLD grown films. To highlight the sharp absorption edge of the CsSnI₃, the absorption coefficient of a reference MAPbI₃ film is presented.*

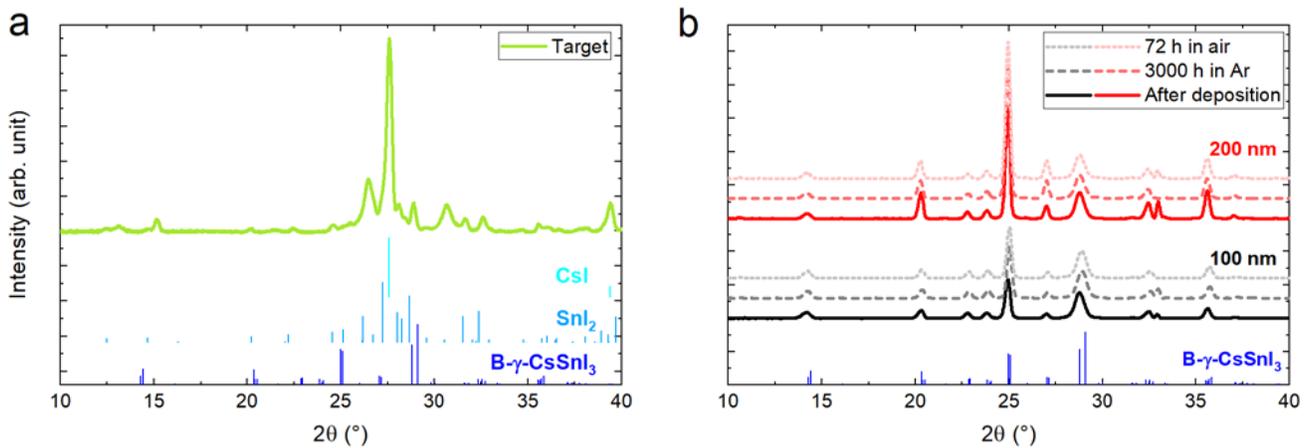

*Figure 3.(a) X-ray diffraction (XRD) pattern of the solid PLD target shown in Figure 1. (b) XRD patterns of 100 and 200 nm thick CsSnI₃ films deposited at room temperature by PLD. The solid lines are measurements directly after the deposition, the dashed lines after 3000 h in Ar atmosphere, and the dotted lines after additional 72 h in air. For comparison, the same B-γ-CsSnI₃ reference spectrum is plotted below in both graphs. The plots indicate that while the source target doesn't present the CsSnI₃ phase but a mixture of the CsI and SnI₂ powders, the resulting thin films present the single black orthorhombic phase of CsSnI₃, which is stable over long time thanks to the Al₂O₃ capping layer.*

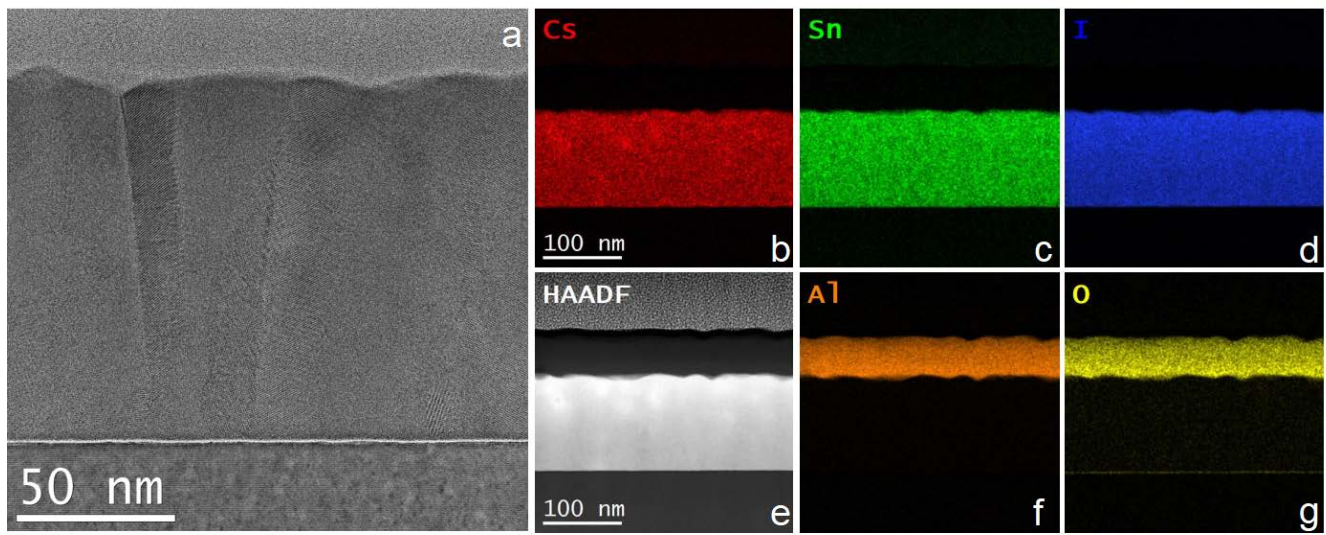

*Figure 4. Cross-section (a) bright-field TEM image of a PLD-grown B-γ-CsSnI$_3$ film on Si. The image shows the formation of a dense film with elongated crystalline grains. (b-g) High-angle annular dark-field (HAADF) image and EDS mapping of the constituent elements of CsSnI$_3$ and the Al$_2$O$_3$ capping layer. EDS confirms uniform distribution of Cs, Sn and I along the thickness of the layer and conformal coating of the Al$_2$O$_3$ capping layer.*

# Table of Contents

High quality films of lead-free halide perovskite caesium tin iodide (CsSnI3) are grown by pulsed laser deposition, a solvent-free, single-source in-vacuum technique. The optically active black γ-CsSnI3 perovskite phase is stabilized through in situ application of an amorphous, transparent capping layer.

*Vivien M. Kiyek, Yorick A. Birkhölzer, Yury Smirnov, Martin Ledinsky, Zdenek Remes, Jamo Momand, Bart J. Kooi, Gertjan Koster, Guus Rijnders, Monica Morales-Masis\**

**TOC figure**

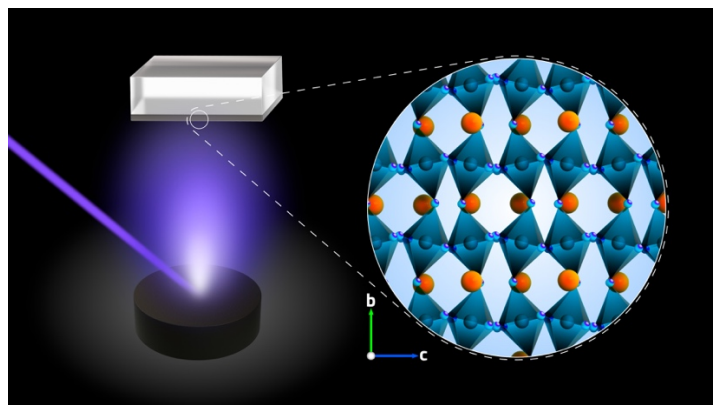

# Supporting Information

# Single-source, solvent-free, room temperature deposition of black γ-CsSnI$_3$ films

*Vivien M. Kiyek, Yorick A. Birkhölzer, Yury Smirnov, Martin Ledinsky, Zdenek Remes, Jamo Momand, Bart J. Kooi, Gertjan Koster, Guus Rijnders, Monica Morales-Masis\**

Pressed Cs-Sn-I target properties

Figure S1 shows photographs of the three fabricated targets. All targets are fabricated following the procedure of Fig.1 (main manuscript) with the only difference on the mixing time of the powders. The powders were mixed in the cylindrical vessel for 3 h (target t$_{mix}$= 3h), 7h (target t$_{mix}$= 7h) and 84 h (target t$_{mix}$= 84h). As observed in the pictures, the color of the targets implies improved uniformity with mixing time. For a more quantitative analysis, the targets were analyzed by energy-dispersive X-ray spectroscopy (EDS). The ratios of Cs, Sn and I measured at 3 different spots of the targets are indicated in Table 1. The obtained ratios demonstrate the improved uniformity of elemental distribution in the target mixed for 84 h. The measured rations are also in accordance with that expected for CsI+SnI$_2$ equimolar powders. The EDS data is normalized to Cs in all cases. We point out that a ratio of 1/1/3 does not explicitly mean the formation of CsSnI$_3$ black perovskite phase. Moreover, oxygen was also detected in the EDS spectra, as expected due to the target exposure in ambient air prior to analysis.

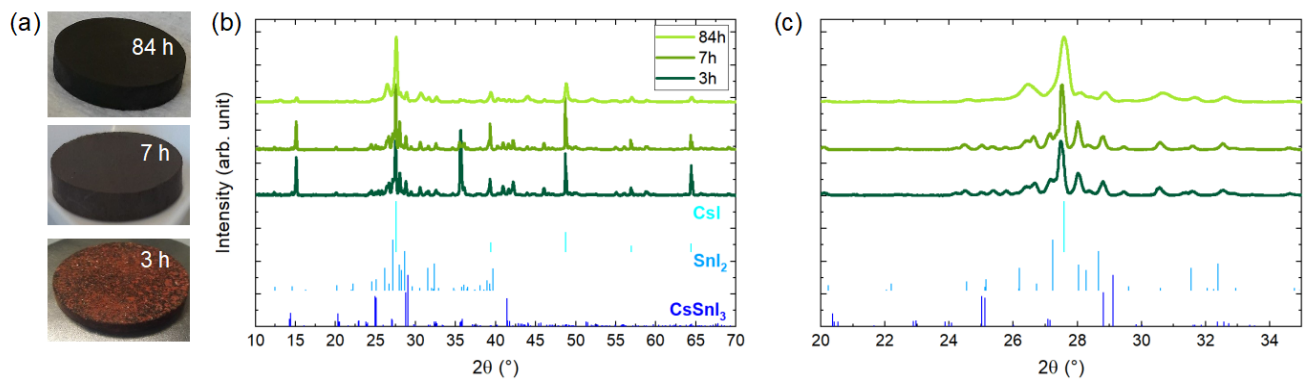

*Figure S1. (a)Photographs of three targets fabricated following the procedure of Fig.1 but with distinct milling times (t$_{mix}$). From top to bottom, t$_{mix}$ = 84h, 7h and 3h. The photographs indicate that increasing the milling time results in an improvement of the elemental distribution in the bulk target. All targets are 2 cm in diameter. (b) XRD data of the three targets synthesized with different mixing times. (c) Zoom-in of feature-rich low-angle range. All major peaks can be assigned to the starting compounds CsI and SnI$_2$ and thus we conclude that no significant chemical reaction took place during ball milling.*

*Table 1. Comparison of target properties including powder mixing time, target density and Cs/Sn/I ratio at three different spots of the targets.*

| Target | $t_{mix}$ = 3h | | | $t_{mix}$ = 7h | | | $t_{mix}$ = 84h | | |
|---|---|---|---|---|---|---|---|---|---|
| Milling (cylindrical vessel, ZrO₂ balls) | 3 h | | | 7 h | | | 84 h | | |
| Density [g/cm³] | 4.61 | | | 4.14 | | | 4.19 | | |
| Cs/Sn/I ratios (normalized to Cs) determined by EDX at different target areas. | Cs | Sn | I | Cs | Sn | I | Cs | Sn | I |
| | 1 | 0.9 | 3 | 1 | 1.6 | 3.6 | 1 | 1 | 2.8 |
| | 1 | 0.5 | 2 | 1 | 1.4 | 3.3 | 1 | 0.9 | 2.8 |
| | 1* | 19* | 36* | 1 | 1.7 | 3.9 | 1 | 0.8 | 2.6 |

* On the target $t_{mix}$=3h (low mixing time target), some highly Cs depleted areas were found. This indicates that also Sn depleted areas must be present but were not measured by EDX.

X-ray diffraction patterns and photoluminescence spectra of films deposited from the three different solid targets

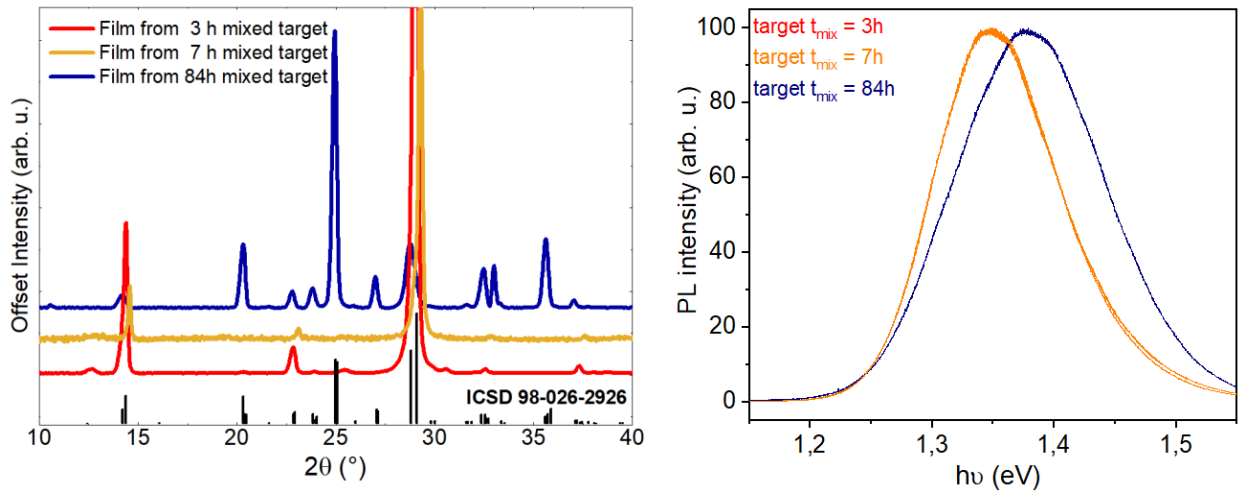

*Figure S2. Left. Comparison of 100 nm-thick CsSnI3 films deposited from two different targets, highlighting the distinct preferential orientations of the films. Right. Steady state photoluminescence (PL) spectra of films grown from the different targets.*

Optical band gap determination

Optical bandgap ($E_g$) of halide absorbers was estimated from the Tauc relation: $\alpha = (h\nu - E_g)^x$ with $\alpha$ being the absorption coefficient obtained from photothermal deflection spectroscopy (PDS) measurement and $h\nu$ being the photon energy. The power factor of the exponent $x$ depends on the nature of absorption transitions. Since $CsSnI_3$ is reported to be a p-type direct band gap semiconductor, we used the exponent $x = 1/2$ corresponding to allowed direct optical transitions. $E_g$ value of $CsSnI_3$ film determined from the intercept of $\alpha^2 = 0$ is 1.32 eV while for the most common hybrid halide perovskite absorber $MAPbI_3$ it is found to be 1.6 eV (Figure S3).

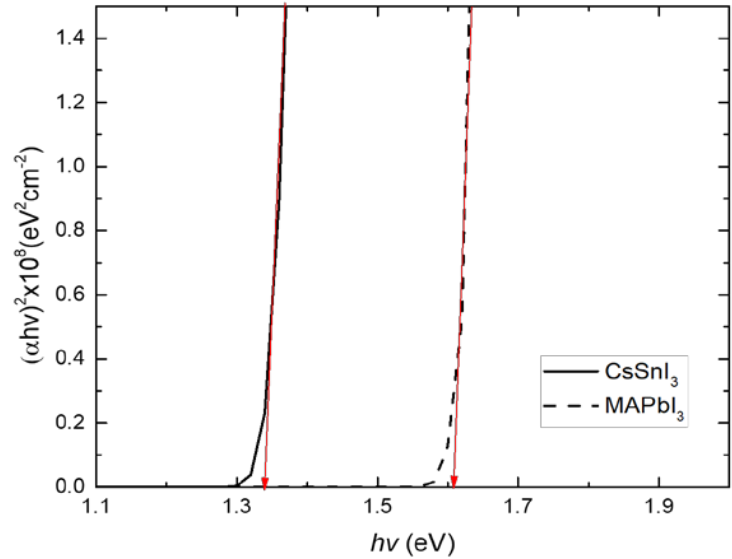

*Figure S3. Optical bandgap for thin films of $CsSnI_3$ (solid line) and $MAPbI_3$ (dashed line) as determined from Tauc plot; absorption coefficient obtained by PDS measurements.*

Rutherford Backscattering Spectroscopy (RBS) and Particle Induced X-ray Emission (PIXE)

RBS and PIXE spectroscopy were performed on the $CsSnI_3$ films. As the films were transported in non-vacuum atmosphere, Sn in the films oxidized ($Sn^{2+} + O^{2-} \rightarrow Sn^{4+}$) following the reaction:

$$2\ CsSnI_3 + O_2 \rightarrow Cs_2SnI_6 + SnO_2$$

The formation of $Cs_2SnI_6$ is confirmed in the RBS spectra. The $SnO_2$ was not possible to quantify due to the low sensitivity to oxygen at 2 MeV RBS, and the presence of the $Al_2O_3$ film on top. As in the reaction from $CsSnI_3$ to $Cs_2SnI_6$, the amount of iodide does not change, we conclude that the $CsSnI_3$ films have a content of ~65 at% of I.

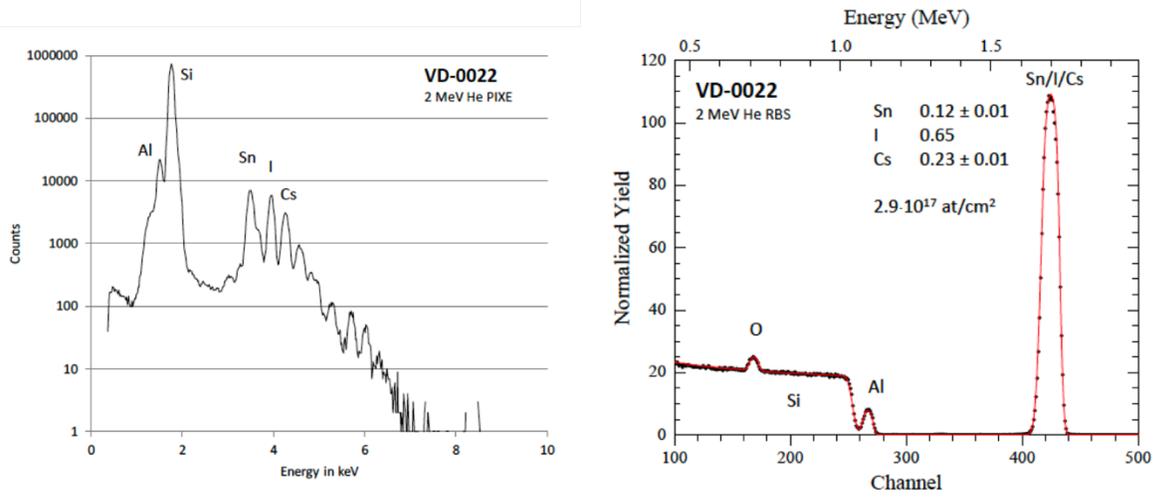

*Figure S4. PIXE (left) and RBS (right) spectra of CsSnI$_3$ films after exposure to air, resulting in Cs$_2$SnI$_6$ from oxidation of Sn (Sn$^{2+}$ + O$^{2-}$ -> Sn$^{4+}$). Samples were measured by 2 MeV He RBS and 2 MeV He PIXE. Solid red lines in RBS spectra are RUMP simulations. PIXE spectra were analyzed by the GUPIX software. Sn/I/Cs atomic ratios were determined by PIXE and are approximately 2/1/6.*

TEM cross-section of the Al$_2$O$_3$ capped CsSnI$_3$ films

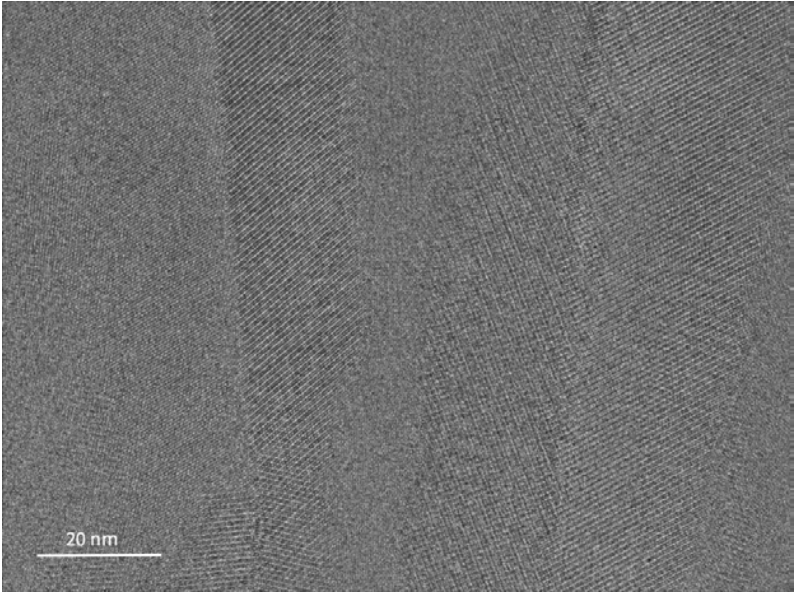

*Figure S5. Zoomed-in area of Fig.4a with enhanced contrast. The atomic planes of the grains forming the polycrystalline CsSnI$_3$ films are observed.*

Surface morphology of the Al₂O₃ capped CsSnI₃ films

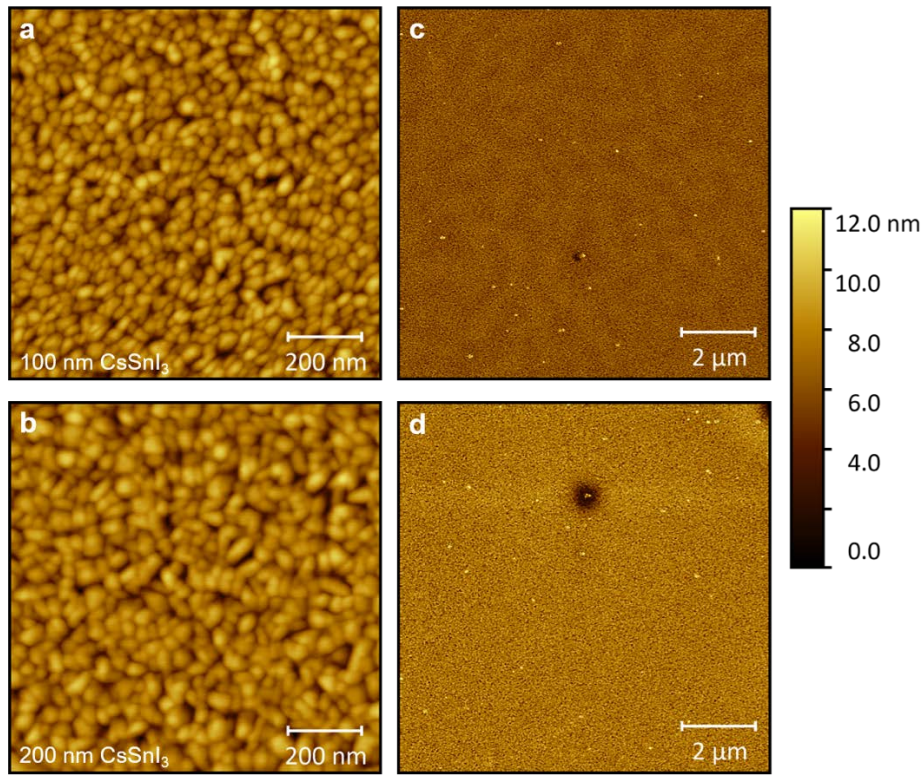

*Figure S6. AFM topography images of 100 nm (a,c) and 200 nm (b,d) CsSnI₃ films with Al₂O₃ capping layer. The measured rms (sq) roughness extracted in Gwyddion is 1.1 nm and 1.2 nm for the 100 (c) and 200 nm (d) thin films, respectively.*

The AFM images were recorded on a Bruker Dimension Icon instrument in tapping mode using Bruker TESPA-V2 cantilevers (nominal tip radius 7 nm and spring constant 42 N/m). The thickness of the Al₂O₃ capping layer was 13 and 40 nm for the 100 and 200 nm thick CsSnI₃ films, respectively.
Surface morphology of the $Al_2O_3$ capped $CsSnI_3$ films

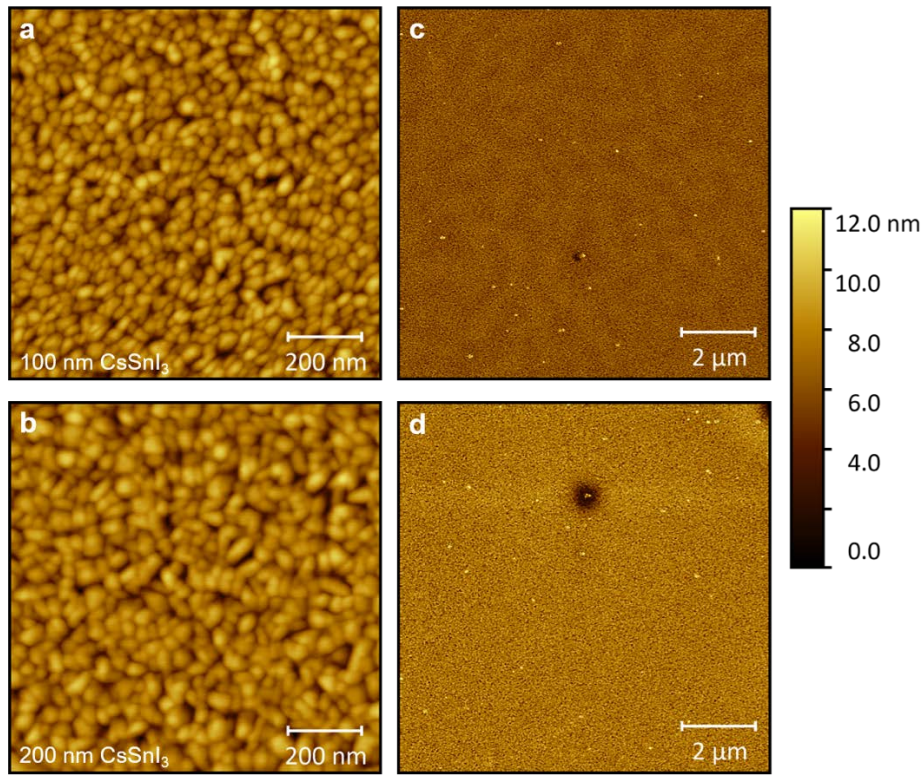

*Figure S6. AFM topography images of 100 nm (a,c) and 200 nm (b,d) $CsSnI_3$ films with $Al_2O_3$ capping layer. The measured rms (sq) roughness extracted in Gwyddion is 1.1 nm and 1.2 nm for the 100 (c) and 200 nm (d) thin films, respectively.*

The AFM images were recorded on a Bruker Dimension Icon instrument in tapping mode using Bruker TESPA-V2 cantilevers (nominal tip radius 7 nm and spring constant 42 N/m). The thickness of the $Al_2O_3$ capping layer was 13 and 40 nm for the 100 and 200 nm thick $CsSnI_3$ films, respectively.

2424Surface morphology of the $Al_2O_3$ capped $CsSnI_3$ films

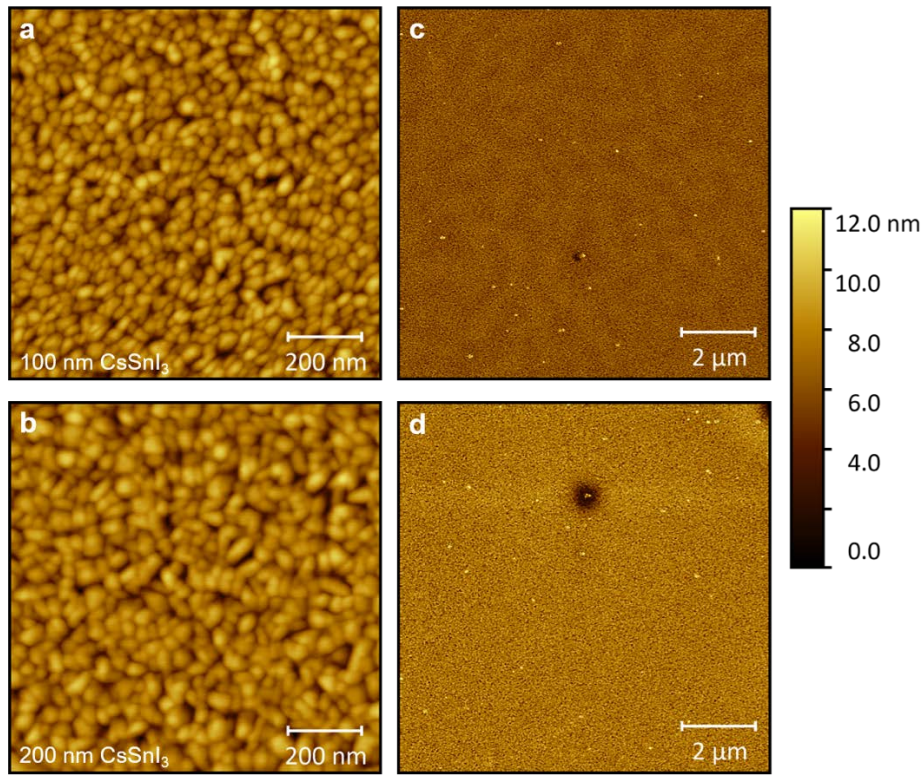

*Figure S6. AFM topography images of 100 nm (a,c) and 200 nm (b,d) $CsSnI_3$ films with $Al_2O_3$ capping layer. The measured rms (sq) roughness extracted in Gwyddion is 1.1 nm and 1.2 nm for the 100 (c) and 200 nm (d) thin films, respectively.*

The AFM images were recorded on a Bruker Dimension Icon instrument in tapping mode using Bruker TESPA-V2 cantilevers (nominal tip radius 7 nm and spring constant 42 N/m). The thickness of the $Al_2O_3$ capping layer was 13 and 40 nm for the 100 and 200 nm thick $CsSnI_3$ films, respectively.

24